%% file: Template.tex
\documentclass{article}
\usepackage[subtle]{savetrees}
\usepackage{spconf,amsmath,graphicx}
\usepackage{subcaption}
\usepackage{amssymb}
\usepackage{lipsum}
\usepackage{xpatch,xcolor}
\usepackage{tabularx}
\usepackage{slashbox}
\usepackage{makecell, booktabs}
\newcolumntype{Y}{>{\centering\arraybackslash}X}
\usepackage{xpatch,xcolor}
\usepackage{float}
\usepackage{multicol,multirow}
\usepackage{url,cite}
\usepackage{hyperref}
\hypersetup{
    colorlinks=true,
    linkcolor=purple,
    filecolor=magenta,      
    urlcolor=purple,
}

\def\BigRoman{\uppercase\expandafter{\romannumeral\number\count 255}}
\def\Romannumeral{\afterassignment\BigRoman\count255=}

\newcommand{\newpara}[1]{\vspace{10pt}\noindent\textbf{#1}}

\title{Advancing the dimensionality reduction of speaker embeddings for speaker diarisation: disentangling noise and informing speech activity}

\name{\parbox{\linewidth}{\centering You Jin Kim$^{1,*}$\thanks{$^{*}$These authors contributed equally to this work.}, Hee-Soo Heo$^{1,*}$, Jee-weon Jung$^{1,*}$, Youngki Kwon$^1$, Bong-Jin Lee$^1$, Joon Son Chung$^2$}}
\address{
  $^1$Naver Corporation, South Korea \\
  $^2$Korea Advanced Institute of Science and Technology, South Korea
}

\begin{document}
\ninept
\maketitle
\begin{abstract}

The objective of this work is to train noise-robust speaker embeddings adapted for speaker diarisation.
Speaker embeddings play a crucial role in the performance of diarisation systems, but they often capture spurious information such as noise, adversely affecting performance. 
Our previous work has proposed an auto-encoder-based dimensionality reduction module to help remove the redundant information. 
However, they do not explicitly separate such information and have also been found to be sensitive to hyper-parameter values.
To this end, we propose two contributions to overcome these issues:
(i) a novel dimensionality reduction framework that can disentangle spurious information from the speaker embeddings;
(ii) the use of speech activity vector to prevent the speaker code from representing the background noise.
Through a range of experiments conducted on four datasets, our approach consistently demonstrates the state-of-the-art performance among models without system fusion.

\end{abstract}
\section{Introduction}
\label{sec:intro}

Speaker diarisation is an interesting but challenging problem. 
The ability to determine ``who spoke when'' provides important context in speech transcription tasks, such as meeting transcription and video subtitling. 
One of the main challenges in speaker diarisation involves the task of clustering speech into an unknown number of speakers. The difficulty is augmented by the challenging environmental characteristics, such as background noise. 

There are two main approaches to solve this challenging problem in previous literature: conventional module-based~\cite{garcia2017speaker, huang2020speaker} and end-to-end~\cite{xue2021online, fujita2020end}. 
The former ``divides-and-conquers'' speaker diarisation into several sub-tasks. The exact configuration differs from system to system, but in general they consist of speech activity detection (SAD), embedding extraction and clustering~\cite{landini2020but, kwon2021look}.
The latter directly segments audio recordings into homogeneous speaker regions using deep neural networks~\cite{horiguchi2020end, kinoshita2021advances}. 
However, current end-to-end approaches have been reported to be strongly overfitted to the environments that they are trained for, not generalised to diverse real-world conditions. Therefore, the winning entries to recent diarisation challenges~\cite{ryant2020third, nagrani2020voxsrc} either exploit the former approach or fuse both approaches.

The performance of the conventional speaker diarisation system which consists of multiple modules, is highly dependent on the ability to cluster the speaker embedding.
Our recent work has proposed a number of methods to adapt the speaker embedding for speaker diarisation~\cite{kwon2021adapting}.
Among such proposals, the dimensionality reduction (DR) module utilised an auto-encoder (AE) trained in an unsupervised manner, and projected speaker embeddings to a low-dimensional code (e.g., $256$ to $20$), adapting towards each session. 
Speaker embeddings in diarisation tasks are only required to discriminate a small number of speakers, compared to thousands in case of verification. Therefore, finding a low-dimensional latent space effectively reduced unnecessary background noise and showed a potential in this line of research.

However, we empirically found that the effectiveness of our DR module varies from session to session. 
When the AE is trained independently for each session, we adopt a fixed code dimensionality, whereas we assume that the optimal code dimensionality may differ in each session depending on two factors: (i) the number of speakers and (ii) the duration. If the dimensionality is too small, the information required for speaker diarisation in the code becomes insufficient, resulting in performance degradation.
In contrast, the excessive dimensionality may cause unnecessary information (e.g., background noise) to reside in the code~\cite{huang2020unsupervised}. 
Furthermore, the existing DR module trains the AE without distinction of speech or non-speech, potentially enforcing the projected embedding to also represent background noise as well as speaker identity~\cite{polyak2019attention}. 
The focus of this work will therefore be on mitigating the limitations of the existing DR module, and improving to be less hyper-parameter-dependent.

We propose two additional improvements upon the existing DR module to accomplish the goal.
First, we extend the AE architecture by adding another code whereby the two codes each stand to represent speaker identity ({\em ``speaker code''}) and other irrelevant information ({\em ``noise code''}), respectively (Section~\ref{ssec:disentangle}). 
Employing two codes, the proposed method excludes noise-relevant factors from the speaker code. 
Second, we introduce {\em ``speech activity vector (SAV)''} to the DR module which represents whether the input is extracted from a speech or a non-speech segment (Section~\ref{ssec:indicator}). 
Training with SAV would ideally force the speaker code to be empty for speaker embeddings from non-speech segments, and therefore prevent the speaker code from representing the background noise.

We evaluate the effectiveness of the proposed methods on a range of datasets, on which we show the state-of-the-art performance (Section~\ref{sec:experiment}). 
In addition, we present additional analysis that our proposed approaches result in a less hyper-parameter-dependent module (Section~\ref{subsec:discussion_2}).

\begin{figure}[t!]
  \centering
  \includegraphics[width=1.0\linewidth, trim={0 10 10 0}, clip]{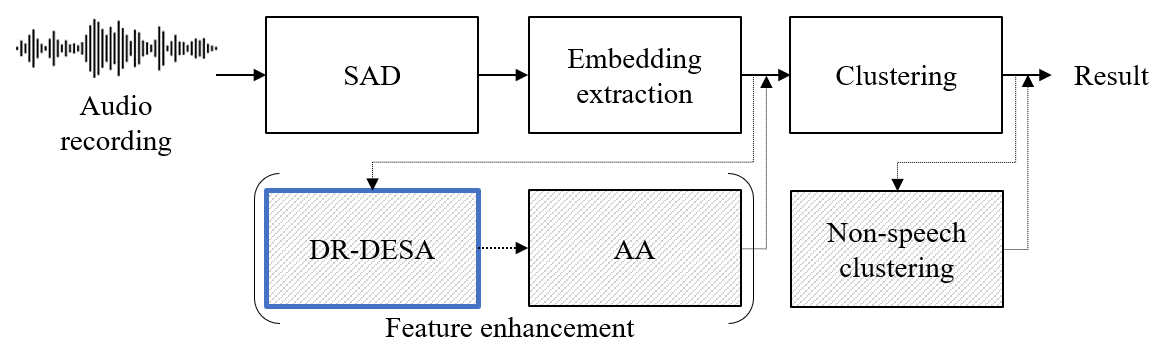}
  \caption{Our speaker diarisation pipeline. Abbreviations are: speech activity detection (SAD), DR (dimensionality reduction), DR-DESA (DR with disentanglement and speech activity vector), attention and aggregation (AA). SAD can be either reference or system one, and DR-DESA is newly proposed in this paper, improving DR~\cite{kwon2021adapting}.} 
  \label{fig:framework}
\end{figure}

\section{Speaker Diarisation Pipeline}
\label{ssec:spekaerDiarisationPipleine}

In this section, we introduce the overall pipeline of our speaker diarisation system, which consists of SAD, speaker embedding extraction, feature enhancement, and clustering modules. 
We omit explanation of SAD because the scope of this work only includes the scenario with a reference SAD. However, our framework can be also applied to system SAD as well.
Figure~\ref{fig:framework} summarises the overall pipeline of our system.


\subsection{Speaker embedding extraction}
For every segment, we extract fixed-dimensional speaker embeddings to represent speaker characteristics from the segments. 
Our speaker embedding extraction module is identical to that of our previous work~\cite{kwon2021adapting}.
It extracts frame-level features using a residual applied trunk network followed by an average pooling layer. 
Each speaker embedding is extracted from a fixed duration of $1.5$ seconds using a sliding window with $0.5$ seconds shift.
The embedding extractor is trained using VoxCeleb1~\cite{nagrani2017voxceleb}, VoxCeleb2~\cite{chung2018voxceleb2}, and MLS~\cite{pratap2020mls} datasets. 
See Section 2.4 of~\cite{kwon2021adapting} for full details.

\subsection{Speaker embedding enhancement}
Our pipeline employs two modules to adapt speaker embeddings that were originally trained for speaker verification towards diarisation: (i) dimensionality reduction with disentanglement and speech activity vector (DR-DESA, addressed in Section~\ref{sec:proposedMethod}); and (ii) attention-based aggregation (AA)~\cite{kwon2021adapting}.
The DR-DESA module refers to the proposed module, which replaces the DR ~\cite{kwon2021adapting} module. 
The DR-DESA and the DR module share the following properties. 
They use a lightweight AE trained for each session. The AE comprises of two fully-connected layers, one for the encoder and the other for the decoder. For the encoder layer, we apply the maximum feature map~\cite{wu2018light} as a non-linearity, whereas the decoder does not adopt one. The differences and the improvements of DR-DESA compared to DR are further described in Section~\ref{sec:proposedMethod}. 

The AA module further refines dimensionality-reduced speaker embedding using a self-attention mechanism.
The module encourages the features located close in the latent space to lie even more closer together, while further pushing distant features apart. 
The objective of this module is to remove noises and outliers on the affinity matrix, using the global context of each session.

\subsection{Clustering}
We assign a speaker label to each speaker embedding using a spectral clustering algorithm~\cite{ning2006spectral} that is widely adopted in the speaker diarisation literature.
We apply eigen-decomposition to speaker embeddings after the DR-DESA and the AA modules without further refinement processes~\cite{ning2006spectral, wang2018speaker} that are typically adopted in existing works.
The number of clusters ({\em i.e.} speakers) is decided by counting the number of eigen-values higher than a predefined threshold; eigen-vectors corresponding to selected eigen-values are used as the spectral embeddings.
Speaker labels are derived using a k-means clustering algorithm on the spectral embeddings. 

\begin{figure}[t]
  \centering
    \begin{subfigure}[t]{0.3\textwidth}
      \includegraphics[width=\textwidth]{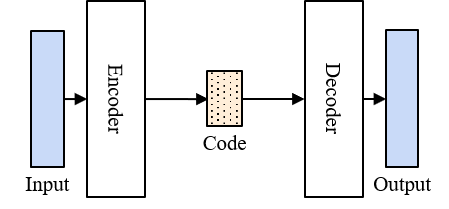}
      \vspace{-15pt}
      \caption{Architecture of DR. The noise may reside in the code of DR. The dots inside the code depicts the nose. }
      \vspace{10pt}
    \end{subfigure}
    \begin{subfigure}[t]{0.35\textwidth}
      \includegraphics[width=\textwidth]{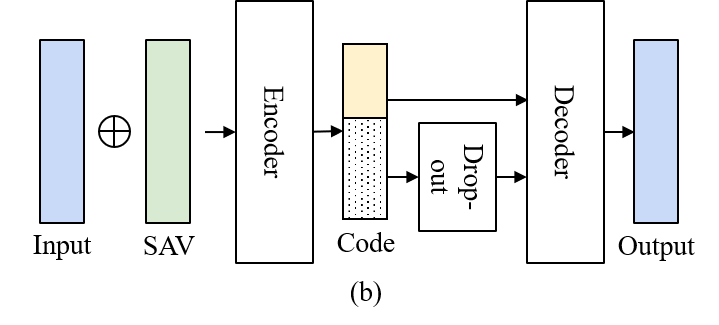}
      \vspace{-17pt}
      \caption{Architecture of the proposed DR-DESA. It disentangles the code into speaker code and the noise code, applying dropout and SAV. The orange part illustrates speaker code, and the dotted part depicts noise code.}
    \end{subfigure} 
  \caption{Architecture comparison between DR~\cite{kwon2021adapting} and the proposed DR-DESA.} 
  \label{fig:dr}
\end{figure}

\section{disentanglement and speech activity vector}
\label{sec:proposedMethod}

We propose a new model referred to as DR-DESA (Figure~\ref{fig:dr}-(b)), extending the original DR (Figure~\ref{fig:dr}-(a)) with two proposals :~(i) we disentangle the existing code by adding noise code and applying dropout to it.  
(ii) we adopt an SAV denoting whether the speaker embedding includes speakers' voice.

\subsection{Embedding disentanglement}
\label{ssec:disentangle}

In the original DR module, an input is projected into a low-dimensional code and then reconstructed.
During this process, the noise factor is inevitably entangled in the code because noise is also required to reconstruct the original input~\cite{chung2019delving}.
The noise factor entangled in the code may disturb speaker clustering, as noise may be consistent across different speakers' identities. 
To mitigate this potential threat, we propose to disentangle the noise factor. 
We divide the latent space into two, and force them to represent speaker-relevant ({\em speaker code}) and irrelevant information ({\em noise code}) respectively. 
We apply dropout only to the noise code, making the neglectful information flow to it.
This is a frequently used technique for disentanglement~\cite{jaiswal2018unsupervised}, where it has been reported that it makes essential information for reconstruction to be gathered in the code where dropout does not exist.  
We concatenate the two codes and feed it to the decoder.
After the training is complete, only the speaker code is used for subsequent clustering step, discarding the noise code. 




\subsection{Speech activity vector}
\label{ssec:indicator}

Using two kinds of codes opens a new potential by discarding speaker-irrelevant information from the speaker code. 
However, the behaviour of the AE becomes more complicated. The speaker code should primarily represent input embeddings extracted from speech segments. 
On the other hand, the noise code should mainly represent input embeddings from non-speech segments. 
To enable this ideal scenario, the AE is required to distinguish whether an input is from speech. 

We further propose to adopt an SAV, which takes the form of a learnable vector which has a dimensionality identical to the input embedding to compose the proposed DR-DESA. 
An SAV is added element-wisely to the input embedding, similar to the positional encoding~\cite{vaswani2017attention}.
Concretely, we adopt two SAVs, one for the speech embedding and the other for the non-speech embedding.
Depending on the speaker embedding's type, we add either SAV to the speaker embedding.
Note that since the SAD is already included in the speaker diarisation pipeline (either system or reference) and precedes the speaker embedding extraction step, we can utilise SAD results at no additional cost.



\section{Experiments}
\label{sec:experiment}

We evaluate the effectiveness of the proposed methods on DIHARD and VoxConverse datasets. The datasets and the experimental details are described in the following paragraphs.

\subsection{Datasets}
\label{ssec:datasets}

\newpara{DIHARD datasets.} The DIHARD challenges publish evaluation datasets which include sessions recorded in restaurant, clinical interview, YouTube videos, etc., making the scenario more challenging. 
We use the evaluation sets of the first, second, and third DIHARD challenges~\cite{ryant2018first, ryant2019second, ryant2020third}. 

\newpara{VoxConverse.} It is an audio-visual speaker diarisation dataset, which consists of speech clips extracted from YouTube videos. The corpus contains overlapped speech, a large speaker pool, and diverse background conditions, including talk-shows, panel discussions, political debates and celebrity interviews~\cite{chung2020spot}.
Test set version {\tt 0.0.2} is used for experiments.

\subsection{Evaluation protocol}
\label{ssec:evaluationProtocol}

Diarisation error rate (DER), the summation of false alarm (FA), missed speech (MS), and speaker confusion (SC), is used as the primary metric. 
FA and MS are related to the SAD module, whereas SC to the DR or the proposed DR-DESA modules.
For all experiments conducted on four datasets, we use the reference SAD to precisely compare the impact of SC caused by either the DR or the proposed DR-DESA.

We use the d-score toolkit\footnote{https://github.com/nryant/dscore} for measuring the DER.
We do not use forgiveness collar for experiments involving the DIHARD datasets, whereas we set a $0.25$ seconds forgiveness collar for VoxConverse experiments to match the scenario with corresponding challenges.

\begin{figure*}[ht!]
    \centering
    \begin{subfigure}[t]{0.275\textwidth}
      \includegraphics[width=\textwidth]{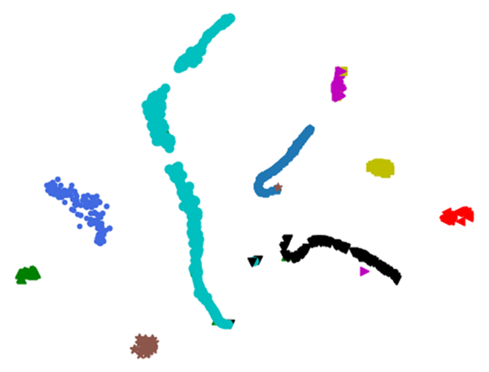}
      \caption{Code generated by DR.}
    \end{subfigure} 
    \hspace{20pt}
    \begin{subfigure}[t]{0.275\textwidth}
      \includegraphics[width=\textwidth]{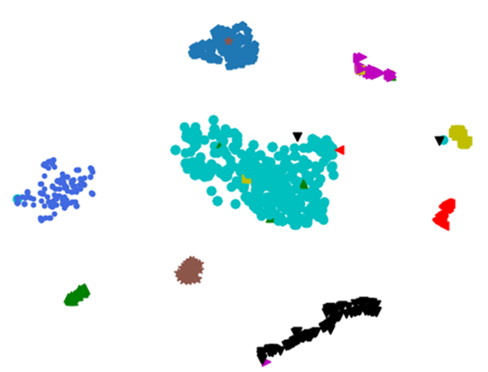}
      \caption{Speaker code generated by DR-DESA.}
    \end{subfigure}   
    \hspace{20pt}
    \begin{subfigure}[t]{0.275\textwidth}
      \includegraphics[width=\textwidth]{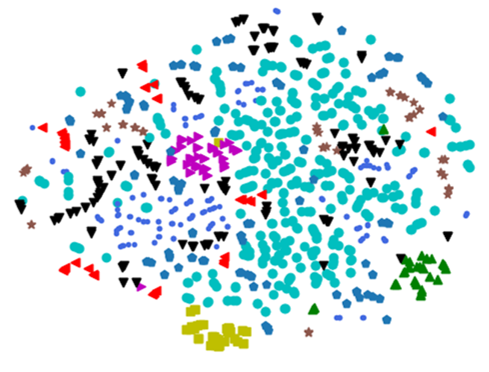}
      \caption{Noise code generated by DR-DESA.}
    \end{subfigure}        
      \caption{Visualisation of the code. Input audio involves nine speakers, resulting in nine clusters ideally. The number of clusters in (a) exceeds nine (the cyan-rounds are divided into several clusters), whereas (b) includes precisely nine clusters. In addition, noise code (c) does not form meaningful clusters.} 
  \label{fig:vis}
\end{figure*}

\subsection{Results}
\label{ssec:resultsOnEachDataset}

Table~\ref{tab:combined_exp} presents the performances of the proposed methods on the four datasets compared with the baselines. 
We also conduct ablation studies where we exclude each proposed component to verify the effect of each component on the overall performance.
Note that, since we utilise reference SAD results, FA is zero in all cases and MS corresponds to the proportion of the overlapped speech included in each dataset. 


\newpara{Comparison with the baselines.}
In all datasets, DR-DESA outperforms the baselines without DR module by a large margin. 
In the case of the DIHARD datasets, the SC error is more than halved, and in VoxConverse SC reduced by more than $30$\%.
In all four datasets, DR-DESA performs even better than the DR consistently. 

\newpara{Comparison with state-of-the-art systems.}
Experimental results on DIHARD {\Romannumeral 1} and {\Romannumeral 2} show that the proposed DR-DESA outperforms the winning systems of the challenges.
DR-DESA also outperforms the best single system in DIHARD {\Romannumeral 3} challenge.
In case of VoxConverse, the test set used in VoxSRC challenge ~\cite{nagrani2020voxsrc} has been recently updated. 
Also, the majority of recent researches apply a system SAD in place of a reference SAD; the VoxSRC challenge which uses VoxConverse only has scenarios that use a system SAD.
Therefore, we did not compare DR-DESA's performance with the systems submitted to the challenge. 

\newpara{Ablation studies.}
DR-DESA has two components on top of the baseline with DR, that are disentanglement and SAV. 
We perform ablation studies by excluding each component from the DR-DESA, and show how each proposal affects the performance. 
In all four datasets, removing disentanglement have a greater impact on the performance. 
However, adopting SAV also consistently improves the performance compared to the baseline with DR.
It is DR-DESA that shows the best performance, and the tendency of the performance gain by each component is consistent across all datasets, signifying that the two proposed techniques are complementary.

\input{tables/combined_exp} 

\begin{figure}[!b]
  \centering
    \begin{subfigure}[t]{0.825\linewidth}
      \includegraphics[width=\textwidth]{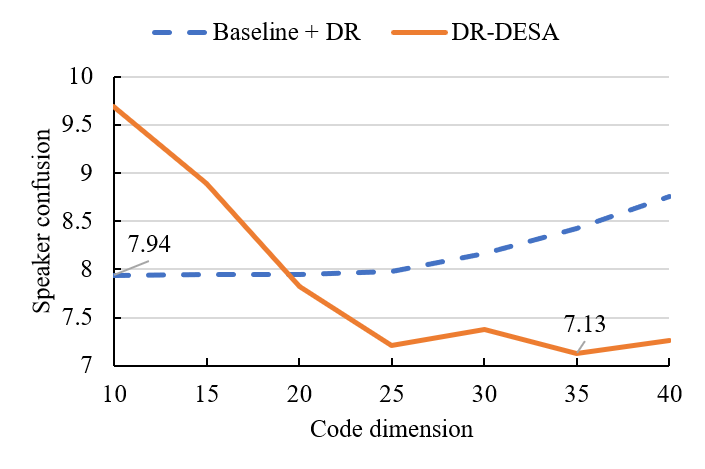}
      \vspace{-20pt}
      \caption{All sessions.}
    \end{subfigure}
    \begin{subfigure}[t]{0.825\linewidth}
      \includegraphics[width=\textwidth]{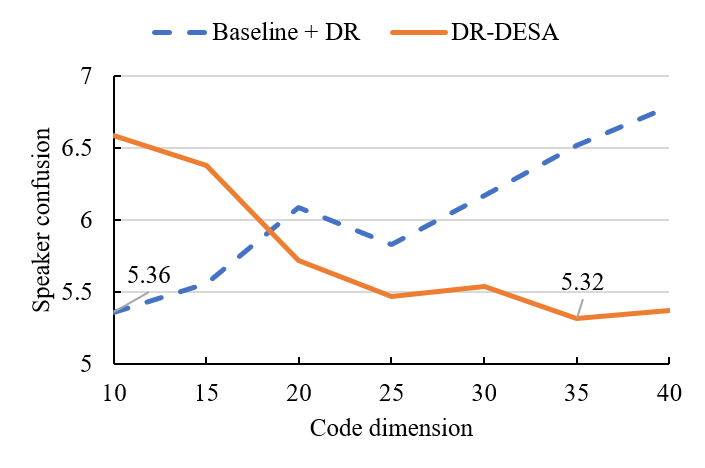}
      \vspace{-20pt}
      \caption{Sessions with four or fewer speakers.}
    \end{subfigure}
    \begin{subfigure}[t]{0.825\linewidth}
      \includegraphics[width=\textwidth]{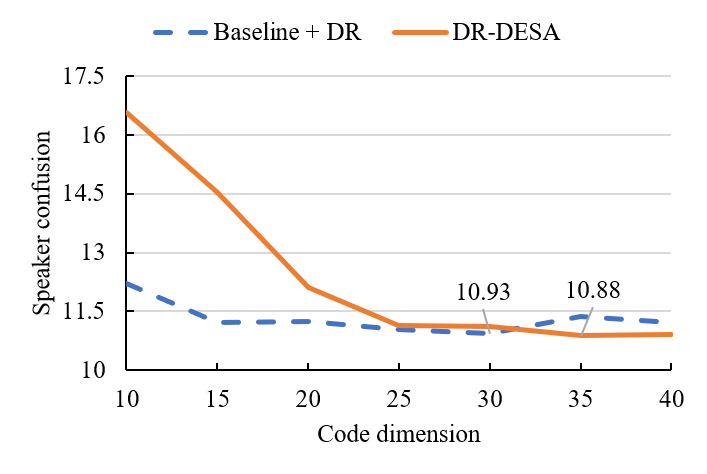}
      \vspace{-20pt}
      \caption{Sessions with five or more speakers.}
    \end{subfigure}    
  \caption{Stability of DR-DESA with high code dimension. (The sessions in DIHARD {\Romannumeral 1}, {\Romannumeral 2}, and {\Romannumeral 3} datasets are used to draw each graph.)} 
  \label{fig:graphs} 
\end{figure}

\section{Further analysis}
\label{sec:discussion}

In this section, we present further analyses to show the role of each code and the strength of DR-DESA. 

\subsection{Visualisation}
Figure~\ref{fig:vis} depicts the code representation of the DR and the DR-DESA module.
Figure~\ref{fig:vis} (a) shows the code from the DR. 
Figure~\ref{fig:vis} (b) represents the speaker code and (c) shows the noise code of the proposed DR-DESA.   
We randomly select an audio recording with nine speakers from the DIHARD {\Romannumeral 2} dataset, extract codes from the audio, and visualise them using t-SNE technique~\cite{van2008visualizing}. 

As shown in the figure, the proposed speaker code (b) represents nine clusters corresponding to nine speakers.
On the other hand, the original code (a) shows more than nine clusters, with the codes of the most dominant speakers divided into multiple clusters. We interpret that this unexpected result is due to the change of noise information within the same speaker, and in the case of the proposed method, this additional information is represented by noise code in (c). This role of the noise code makes the speaker code in (b) have more suitable distribution for speaker diarisation.

\subsection{Analysis based on the number of speakers}
\label{subsec:discussion_2}
We present Figure~\ref{fig:graphs} to show the limitation of DR module and the effectiveness of our DR-DESA using the three DIHARD datasets. 
We evaluate the performance of our baseline (DR module of \cite{kwon2021adapting}) and the proposed DR-DESA using diverse code dimensionalities. 
(a) shows SC of the entire sessions, (b) indicates SC of the sessions where the number of speakers is four or fewer, and (c) shows SC of the session with more than four speakers. 
As argued, the baseline requires low dimensionality for sessions with fewer speakers, and high dimensionality for sessions with more speakers. 
Performance degradation is observed, especially in (b), when the dimensionality is not ideal.
In contrary, our proposed DR-DESA module demonstrates the stable and optimal performance regardless of the number of speakers, when dimensionality is $30$ or more. 
As a result, this stability leads relatively higher performance improvements in the entire dataset, even though the optimal performances of the two systems in each subset do not show a significant difference.

\section{Conclusion}
\label{sec:conclusion}
This paper addresses a novel unsupervised disentanglement framework, which generates noise-robust speaker embeddings for speaker diarisation. 
Speaker embeddings are the crucial component of diarisation systems, but they often contain the unnecessary information that degrades the performance, such as background noise and reverberation.
Recently proposed DR module reduces the dimensionality of the embeddings, in order to remove the spurious information.
However, the effect of DR is limited, being sensitive to the code dimensionality.   

To this end, we propose DR-DESA introducing two more techniques on top of the DR module: (i) explicit disentanglement of the spurious information from the original code; (ii) the introduction of SAV. 
DR-DESA show the state-of-the-art performance as a single system on four benchmark datasets, and ablation studies on DR-DESA demonstrate that both of the proposals lead to performance gains.
In addition, visualising the disentangled code confirms that DR-DESA performs as intended. 

\clearpage  

\bibliographystyle{IEEEbib}
\bibliography{refs}

\end{document}

%% file: tables/combined_exp.tex
\begin{table}[!t]
	\centering
	\caption{Results on DIHARD {\Romannumeral 1}, {\Romannumeral 2}, {\Romannumeral 3}, and VoxConverse datasets \textit{(DER: diarisation error rate, FA: false alarm, MS: miss, SC: speaker confusion)}. DR stands for dimensionality reduction, and DE for disentanglement, and SAV for speech activity vector. DR-DESA for DR with disentanglement and SAV is proposed method with two improvements (DE and SAV).}
	\begin{tabularx}{\columnwidth}{lYYYY}
    \Xhline{1pt}
	 Configuration & DER & FA & MS & SC\\ 
    \Xhline{1pt}
	\multicolumn{5}{c}{DIHARD {\Romannumeral 1}}\\
	\hline
	 Track 1 winner~\cite{sell2018diarization} & 23.73  & - & - & - \\
	 \hline
	 Baseline & 25.85 & 0.00 & 8.71 & 17.14\\
	 Baseline + DR & 17.70 & 0.00 & 8.71 & 8.98\\
	 \hline
	 DR + DE & 17.04 & 0.00 & 8.71 & 8.33\\	 
	 DR + SAV & 17.25 & 0.00 & 8.71 & 8.54\\	 
	 DR + DE + SAV (DR-DESA) & \textbf{16.75} & 0.00 & 8.71 & \textbf{8.04}\\	 
	\hline
	\multicolumn{5}{c}{DIHARD {\Romannumeral 2}}\\
	\hline    
	 Track 1 winner~\cite{landini2019but} & 18.42  & - & - & - \\
	\hline   
	 Baseline & 27.39 & 0.00 & 9.69 & 17.70\\
	 Baseline + DR & 18.40 & 0.00 & 9.69 & 8.71\\
	 \hline
	 DR + DE & 17.76 & 0.00 & 9.69 & 8.08\\    
	 DR + SAV & 18.21 & 0.00 & 9.69 & 8.52\\    
	 DR + DE + SAV (DR-DESA) & \textbf{17.44} & 0.00 & 9.69 & \textbf{7.75}\\    
	\hline
	\multicolumn{5}{c}{DIHARD {\Romannumeral 3}}\\
	\hline   
	 Track 1 best single system~\cite{alamcrim} & 15.50 & - & - & - \\
	 \hline  
	 Baseline & 20.99 & 0.00 & 9.52 & 11.47\\
	 Baseline + DR & 15.49 & 0.00 & 9.52 & 5.97\\
	 \hline
	 DR + DE & 15.28 & 0.00 & 9.52 & 5.76\\	     
	 DR + SAV &15.32 & 0.00 & 9.52 & 5.80\\	     
	 DR + DE + SAV (DR-DESA)  & \textbf{15.05} & 0.00 & 9.52 & \textbf{5.53}\\	     
	\hline
	\multicolumn{5}{c}{VoxConverse}\\
	\hline  
	 Baseline & 5.83 & 0.00 & 1.60 & 4.23\\
	 Baseline + DR & 4.58 & 0.00 & 1.60 & 2.98\\
	 \hline
	 DR + DE & 4.51 & 0.00 & 1.60 & 2.91\\	 
	 DR + SAV & 4.55 & 0.00 & 1.60 & 2.95\\	
	 DR + DE + SAV (DR-DESA) & \textbf{4.45} & 0.00 & 1.60 & \textbf{2.85}\\	
    \Xhline{1pt}
	\end{tabularx}
	\label{tab:combined_exp}
\end{table}